\documentclass[aps,physrev,preprint,superscriptaddress,showkeys]{revtex4-2}
\usepackage{amsmath}
\usepackage[normalem]{ulem} 
\usepackage{xcolor}
\numberwithin{equation}{section}

\usepackage[normalem]{ulem} 
\usepackage{xcolor}
\allowdisplaybreaks
\begin{document}
\title{Non-vacuum metrics for the Newman-Unti-Tamburino background: A coordinate-free approach to diverging and twisting solutions }
\author{Ayşe Hümeyra Bilge}
\affiliation{Department of Industrial Engineering, Kadir Has University, 34083, Istanbul, Türkiye.}
\author{Tolga Birkandan}
\affiliation{Department of Physics, Istanbul Technical University, 34469, Istanbul, Türkiye.}
\author{Tekin Dereli}
\affiliation{Department of Physics, Koç University, 34450, Istanbul, Türkiye.}
\affiliation{Faculty of Aviation and Aeronautical Sciences, Özyeğin University, 34794, Istanbul, Türkiye.}
\author{Gulay Karakaya}
\email{Contact author: E-mail: gulaykarakaya@itu.edu.tr}
\affiliation{Department of Industrial Engineering, Kadir Has University, 34083, Istanbul, Türkiye.}
\affiliation{Department of Physics, Istanbul Technical University, 34469, Istanbul, Türkiye.}
\date{\today}
\begin{abstract}
The geometry of the Newman-Unti-Tamburino (NUT) vacuum solution is characterized as the unique Petrov Type D vacuum metric such that the two double principal null directions form an integrable distribution. We study expanding and twisting non-vacuum Type D metrics in this geometry,  with the additional assumption $\Phi_{01}=\Phi_{12}=0$. We prove that these conditions determine the solutions up to a freedom in $\Phi_{11}\pm 3\Lambda$. 
\end{abstract}
\keywords{NUT solution, Integrable systems, Newman-Penrose formalism, Petrov Type D metrics.}
\maketitle
\newpage
\section{Introduction}
Einstein's field equations relate the geometry of the spacetime to the energy density and pressure of the matter. In local coordinates $\{x^\mu\}$, $\mu=1,\dots,4$, Einstein's equations without the cosmological constant, have the form
\begin{eqnarray}
	R_{\mu\nu}\!-\!\frac{1}{2}R g_{\mu\nu}\!= \kappa_0T_{\mu\nu},
\end{eqnarray}
where $R_{\mu\nu}$ is the Ricci tensor, $g_{\mu\nu}$ is the metric tensor, $R$ is the Ricci scalar, $T_{\mu\nu}$ is the energy-momentum tensor, and $\kappa_0$ is a constant. These equations form a set of $10$ nonlinear second-order partial differential equations for the components of the metric tensor. Vacuum solutions are characterized by the equations $R_{\mu\nu}=0$ while the study of non-vacuum solutions usually initiates with an appropriate choice of matter fields $T_{\mu\nu}$, supplemented with assumptions on the geometry of the underlying manifold and its symmetries.  

When no physical fields are present, solving Einstein's field equations for vacuum is a purely mathematical problem, consisting essentially of determining the compatibility of various geometric assumptions. Whether an analytic explicit solution is obtained or not, the existence of a unique solution to the set of Einstein's equations for vacuum, supplemented by geometrical constraints, is actually the characterization of a ``physical" spacetime, provided that certain inequalities are also satisfied. As a typical example, we can cite Birkhoff's theorem, stating that the only vacuum solution with spherical symmetry is the Schwarzschild solution \cite{Birkhoff}.

In the present work, we investigate to what extent the geometry of the underlying manifold determines the physical fields that are compatible with a given geometry. The starting point is the Type D vacuum solution studied by Newman, Unti and Tamburino, also known as the NUT solution \cite{Newman:1963yy}. In a recent paper \cite{Baysazan}, we have shown that this solution is uniquely characterized by the following three conditions: \\
(i) $R_{\mu\nu}=0$,\\
(ii) The Weyl tensor is of Petrov Type D, \\
(iii) The repeated null directions of the Weyl tensor form an integrable distribution. \\

Here,  we investigate sources compatible with this geometry (NUT geometry), replacing the assumption (i) by the assumption 
(i\textsc{\char13}) as below.\\
(i\textsc{\char13}) Subspaces spanned by the  repeated null directions and their orthogonal complement are invariant under the Ricci tensor in the mixed form.\\

This assumption corresponds to $\Phi_{01}=\Phi_{12}=0$ and  we classify these energy-momentum tensors of physical fields compatible with the NUT geometry. Computations are carried out in the framework of the Newman-Penrose (NP) formalism, and integrability conditions are  used to obtain systems in ``involution".

The NP formalism, introduced in the early 1960s, has been an important tool for obtaining exact solutions, especially for analyzing the asymptotic behavior of gravitational fields \cite{Newman:1961qr,Stephani:2003tm}.
The NP formalism is actually a special case of the moving frame approach. The moving frame is composed of four null vector fields, referred to as the null tetrad. The NP equations describe the components of the connection and curvature, with the directional derivative operators being dual to the one-forms of the moving frame. The components of the connection with respect to the tetrad frame are called ``spin coefficients". The NP equations define explicit formulations for the directional derivatives of the spin coefficients and can be regarded as characterizing the curvature. The components of the Weyl and Ricci tensors appear in the NP equations, and explicit expressions of the  Bianchi identities giving directional derivatives of the components of the curvature are part of the NP system of equations. Algebraic classifications of Weyl and trace-free Ricci tensors, known as the Petrov and Segre-Plenbanski classifications, and symmetry assumptions are commonly used for selecting sub-cases of interest in the search of exact solutions. 
The commutation relations of the null tetrad, the NP equations, the Bianchi identities, and the gauge transformations together form a coordinate-free framework for Einstein's equations. In the literature, typically, the null tetrad is constructed from the metric, and coordinate transformations are applied to obtain the final form of the exact solution in local coordinates.
 
Our approach falls in the framework of the Riquier-Janet theory, which deals with overdetermined systems of partial differential equations \cite{Janet,schwarz1984}. The Riquier-Janet existence theorem is an analogue of the Cauchy-Kowalewski theorem for overdetermined systems, stating essentially that an overdetermined system in ``involution" has a local analytic solution where the free functions and free parameters are determined from the structure of the system. In our approach, we aim to impose geometric conditions that lead to a system in involution where the freedom consists of a finite number of constants. In such cases, one may assert the existence of a solution depending on these parameters, without explicitly obtaining the functional form of the metric.

The outline of the paper is as follows. In Sec.\ref{sec2}, we present the NP equations under certain assumptions for shear-free, expanding and twisting non-vacuum Type D metrics. Then, we obtain the integrability conditions of the NP equations. In Sec.\ref{sec3}, we work out the Ricci classification with non-zero $\Phi_{00}$, $\Phi_{11}$ and $\Phi_{22}$. We determine the Segre types for our system and then find the physical solutions corresponding to each Segre type. We introduce the curvature of the connection in the NP formalism. Finally, in Sec.\ref{sec6}, we summarize our conclusions. 

\section{Preliminaries}\label{sec2}
In a previous work \cite{Baysazan}, we have shown that the Newman-Unti-Tamburino (NUT) solution  \cite{Newman:1963yy} is the unique vacuum Type D  metric with the property that (i) the repeated null directions of the Weyl tensor are diverging and twisting and (ii) the corresponding tangent vectors form an integrable distribution. 
As shown in \cite{Baysazan}, this geometric condition uniquely determines the NUT metric as the Type D vacuum solution, up to coordinate transformations of the $2$-dimensional spacelike sub-manifold.

In the present work, as a follow-up of \cite{Baysazan}, we investigate non-vacuum Type D metrics by similar integrability methods. 
\vskip 0.2cm\noindent
{\subsection{Assumptions and gauge transformations}

We study solutions of Einstein's field equations under the following assumptions.
\begin{itemize}
\item The Weyl tensor is of Type D, and the gauge is chosen to set 
\begin{equation}
\Psi_0=\Psi_1=\Psi_3=\Psi_4=0,\quad \Psi_2\ne 0.
\end{equation}
\item The repeated null directions of the Weyl tensor, $l$ and $n$  are geodesic and shear-free, 
\begin{eqnarray}
\kappa=\sigma=0,\quad \nu=\lambda=0.
\end{eqnarray}
\item The principal null directions of the Weyl tensor form an  integrable distribution, leading to 
\begin{equation}
\bar{\tau}+\pi=0.
\end{equation}
\item The subspaces spanned by $l,n$ and $m,\bar m$ are invariant under the trace-free Ricci tensor (in mixed form), hence  
\begin{eqnarray}
\Phi_{01}=\Phi_{12}=0.   
\end{eqnarray}
\item For $\kappa=0$ and 
$(\rho-\bar\rho)(\bar\tau+\pi)=0$,
$SL(2,C)$ rotations of Type B \cite{Baysazan} are used to set 
\begin{eqnarray}
 \epsilon=0, \quad  \tau=\bar\alpha+\beta. 
\end{eqnarray}
\end{itemize}
\vskip 0.2cm\noindent

In the vacuum case, $\bar{\tau}+\pi=0$ implies that $\tau=0$ and this leads to $\bar{\alpha}+\beta=0$.  In the non-vacuum case, this is not true in general; it holds when $\Phi_{01}=\Phi_{12}=0$, corresponding to the case where the Ricci tensor $R^a_b$ is block diagonal. Furthermore, with the assumptions on the spin coefficients, it turns out that $\Phi_{02}=0$.
The eigenvalues of the traceless Ricci tensor are $\Phi_{11}$ as a double root, and $\Phi_{11}\pm \sqrt{\Phi_{00}\Phi_{22}}$.  
\vskip 0.2cm\noindent
{\bf Proposition 1.} Let $\kappa=\nu=0$, $\epsilon=0$   and $\Psi_2\!-\!\bar{\Psi}_2\neq0$. Then, $\bar\tau+\pi=0$ implies $\tau=0$.
\\ \noindent
{\it Proof.} The Newman-Penrose equations under these assumptions 
imply that $D\tau\!=\!0$ and $\Delta\pi\!=\!\left(\bar{\gamma}\!-\!\gamma\right)\pi$. 
Using $\tau\!+\!\bar{\pi}\!=\!0$, we find $\Delta\tau\!=\!\left({\gamma}\!-\!\bar{\gamma}\right)\tau$.
Applying the commutation relation between
$D$ and $\Delta$ to $\tau$, we obtain $\left(\Psi_2\!-\!\overline{\Psi}_2\right)\tau\!=\!0$.
Thus  $\tau\!=\!0$ since  $\Psi_2\!\neq\!\overline{\Psi}_2$.
\vskip 0.2cm\noindent
{\bf Proposition 2.} Let $\kappa=\sigma=\lambda=\pi=0$, 
then $\Phi_{02}=0$.\\
\noindent
{\it Proof.} Follows directly from the NP equations.
\vskip 0.2cm\noindent

\subsection{Newman-Penrose equations}
 Under the above assumptions, the NP quantities that are zero are listed below. In the following, we will assume that $\rho$ and $\mu$ are both complex.
 \begin{eqnarray}
 &&\kappa=\sigma=0,\quad \nu=\lambda=0,\quad \epsilon=0,
 \quad \tau=\pi=0,\quad \bar\alpha+\beta=0,\quad 
 \rho\ne \bar\rho,\quad \mu\ne \bar\mu,\nonumber\\
 &&\Psi_0=\Psi_1=\Psi_3=\Psi_4=0,\quad \Phi_{01}=\Phi_{02}=\Phi_{12}=0.
 \end{eqnarray}
The systems of equations for each of the nonzero variables are as below.
\begin{eqnarray}
&&D\rho=\rho^2 +\Phi_{00},\nonumber\\
&&\Delta\rho=-\rho\bar\mu +(\gamma+\bar\gamma)\rho-\Psi_2-2\Lambda,\nonumber\\
&&\delta\rho=0,\\
&&\nonumber\\
&&D\mu=\bar\rho\mu+\Psi_2+2\Lambda,\nonumber\\
&& \Delta\mu=-\mu^2-\mu(\gamma+\bar\gamma)-\Phi_{22},\nonumber\\
&&\bar\delta\mu=0,\\
&&\nonumber\\
&&D\alpha=\rho\alpha,\nonumber\\
&&\Delta\alpha-\bar\delta\gamma=\alpha(\bar\gamma-\gamma-\bar\mu),\nonumber\\
&&\delta\alpha+\bar\delta\bar\alpha=\mu\rho+4\alpha\bar\alpha +
\gamma(\rho-\bar\rho)-\Psi_2+\Phi_{11}+\Lambda,\\
&&\nonumber\\
&&D\gamma=\Psi_2+\Phi_{11}-\Lambda,\nonumber\\
&&\delta(\gamma+\bar\gamma)=0,\\
&&\nonumber\\
&&D(\Psi_2-\Phi_{11}-\Lambda)=3\rho\Psi_2+\mu\Phi_{00}-2\bar\rho\Phi_{11},\nonumber\\
&&\Delta(\Psi_2-\Phi_{11}-\Lambda)= -3\mu\Psi_2+\rho\Phi_{22}-2\bar\mu\Phi_{11},\\
&&\nonumber\\
&&\delta(\Psi_2+2\Lambda)=0,\nonumber\\
&&\bar\delta(\Psi_2+2\Lambda)=0,\\
&&\nonumber\\
&&\Delta\Phi_{00}+D(\Phi_{11}+3\Lambda)=(2\gamma+2\bar\gamma-\mu-\bar\mu)\Phi_{00}+2(\rho+\bar\rho)\Phi_{11},\nonumber\\
&&\delta\Phi_{00}=\bar\delta\Phi_{00}=0,\\
&&\nonumber\\
&&D\Phi_{22}+\Delta(\Phi_{11}+3\Lambda)=(\rho+\bar\rho)\Phi_{22}-2(\mu+\bar\mu)\Phi_{11},\nonumber\\
&&\bar\delta\Phi_{22}=\delta\Phi_{22}=0,\\
&&\nonumber\\
&&\delta(\Phi_{11}-3\Lambda)=\bar\delta(\Phi_{11}-3\Lambda)=0.
\end{eqnarray}

\vskip 0.2cm\noindent
{\bf Remark  1.} 
The right-hand side of the equation for $\delta\alpha+\bar\delta\bar\alpha$ determines the real part of $\gamma$, provided that $\rho-\bar\rho$ is nonzero, as below.
\begin{eqnarray}
\gamma+\bar\gamma=\frac{1}{\rho-\bar\rho}
\left[-\mu\rho+\bar\mu\bar\rho+\Psi_2-\bar\Psi_2\right].
\end{eqnarray}
Actually, the imaginary part of $\gamma$ will be determined by an $SL(2, C)$ rotation, but to prove this, we need to show that $\mu\bar\rho-\bar\mu\rho=0$.
\vskip 0.2cm\noindent
The algebraic relation defining $\gamma+\bar\gamma$ should satisfy the equations for $D\gamma$ and $\delta(\gamma+\bar\gamma)=0$. $D\gamma$ does not give any constraint, but $\delta(\gamma+\bar\gamma)=\bar\delta(\gamma+\bar\gamma)=0$  give the following equations.
\begin{eqnarray}
&&\delta\mu=\delta\bar\rho\ (\rho-\bar\rho)^{-1}
[\ \ (\Psi_2-\bar\Psi_2)/\rho-(\mu-\bar\mu)],\\
&&\bar\delta\bar \mu=\bar\delta\rho\ (\rho-\bar\rho)^{-1}
[-(\Psi_2-\bar\Psi_2)/\bar\rho+(\mu-\bar\mu)].
\end{eqnarray}


We recall that if $\rho-\bar{\rho}\neq0$ and $\mu-\bar{\mu}\neq0$, the commutation relation of $\delta$ and $\bar\delta$ determines  $D$ and $\Delta$ derivatives. 
\vskip 0.2cm\noindent
{\bf Proposition 3.}
Assume that $\delta\phi=\bar\delta\phi=0$, $D\phi\ne 0$, $\Delta \phi\ne 0$, $\rho-\bar\rho\ne 0$ and $\mu-\bar\mu\ne 0$. Then
$\Delta\phi=-\frac{\mu-\bar\mu}{\rho- \bar\rho}D\phi$,  
 $\delta\Delta\phi=\delta D\phi=0$, and $\bar\delta\Delta\phi=\bar\delta D\phi=0 $.

\noindent
{\it Proof.}
The commutation relation 
$\bar\delta\delta-\delta\bar\delta=
-(\mu-\bar\mu)D-(\rho-\bar\rho)\Delta
-2\bar\alpha\bar\delta+2\alpha\delta$
gives
\begin{eqnarray}
\Delta\phi=-\frac{\mu-\bar\mu}{\rho- \bar\rho}D\phi.    
\end{eqnarray}
Then it can be seen that the commutation relations 
$ \delta\Delta-\Delta\delta=(\mu-\gamma+\bar\gamma)\delta   $
and $
 \delta D- D\delta=\!-\!\bar\rho \delta $
imply   
\begin{eqnarray}
&&\delta\Delta\phi=0,\quad \delta D\phi=0\\
&&\bar\delta\Delta\phi=0,\quad \bar\delta D\phi=0.
\end{eqnarray}    
\vskip0.2cm

Applying  $\delta$ and $\bar\delta$   (II.19), and using (II.20) and (II.21), with $\phi=\Psi_2$ for example, it is easy to see that 
\begin{eqnarray}
&&    \delta\mu (\rho-\bar\rho) + \delta\bar\rho (\mu-\bar\mu)=0,\\
&&    \bar\delta\bar\mu (\rho-\bar\rho) + \bar\delta\rho (\mu-\bar\mu)=0.
\end{eqnarray}
Comparing (II.17-18)  and (II.22-23), and using the fact that $\Psi_2-\bar\Psi_2\ne0$ it can be seen that
\begin{eqnarray}
    \bar\delta\rho=\delta\bar\rho=0,\quad 
 \delta\mu=\bar\delta\bar\mu=0.
 \end{eqnarray}

\subsection{Algebraic relations in the  NP system}
We start by applying the commutator $[\delta,\bar\delta]$ to $\rho$, $\bar\rho$, $\mu$ and $\bar\mu$. As the $D$ and $\Delta$ derivatives of these variables are known, we obtain algebraic relations. In particular, $[\delta,\bar\delta]$ applied to $\rho-\bar\rho$  gives the well-known identity
\begin{eqnarray}
\mu\bar\rho-\bar\mu\rho=0,
\end{eqnarray}
and the definitions of $\Phi_{00}$ and $\Phi_{22}$ as below.
\begin{eqnarray}
&&\Phi_{00}=(\mu-\bar\mu)^{-1}
\left[-(\Psi_2+2\Lambda)\bar\rho+(\bar\Psi_2+2\Lambda)\rho+2\rho\bar\rho(\mu-\bar\mu)\right],\\
&&\Phi_{22}=(\rho-\bar\rho)^{-1}
\left[-(\Psi_2+2\Lambda)\bar\mu+(\bar\Psi_2+2\Lambda)\mu+2\mu\bar\mu(\rho-\bar\rho)\right].
\end{eqnarray}
Furthermore, it has been checked that
\begin{eqnarray}
\Phi_{22}\!=\!
\frac{(\mu\!-\!\bar{\mu})^2 }{(\rho-\bar{\rho})^2}\Phi_{00}=\frac{\mu^2}{\rho^2}\Phi_{00}.
\end{eqnarray}

It has been checked that these expressions above are consistent with previously defined derivatives of $\bar\mu$, $\Phi_{00}$, and $\Phi_{22}$.
We now show that the imaginary part of $\gamma$  can be fixed by an $SL(2,C)$ rotation. 

Under Type B transformations as given in \cite{Carmeli:1975wq}, the null tetrad transforms as
\begin{eqnarray}
l'_\mu=Al_\mu,\quad m'_\mu=e^{i\theta}m_\mu,\quad n'_\mu=A^{-1}n_\mu,    
\end{eqnarray} 
where $A$ and  $\theta$ are real.
Note that, if $z=A^{1/2}e^{i\theta/2}$, and $\partial$ is any directional derivative operator,  $z \partial z^{-1}=
-{\textstyle \frac{1}{2}}A^{-1}\partial A
-{\textstyle \frac{i}{2}}\partial \theta$.
The conditions $\kappa=\nu=\lambda=\sigma=0$ are preserved under these transformations, and $\Psi_2$ is invariant.
Recall that 
$\epsilon$ is set to zero by determining $Dz$. Then for  $\kappa=0$, and $\bar\tau+\pi=0$, one can set 
$\tau=\bar\alpha+\beta$.  Furthermore, we have shown that if $\Phi_{01}=\Phi_{12}=0$,  $\bar\tau+\pi=0$ implies $\tau=\pi=0$, and this, in turn gives $\bar\alpha+\beta=0$.  As $\rho-\bar\rho$ is nonzero, this determines the real part of $\gamma$. Under these conditions, the imaginary part of $\gamma$  is determined by an $SL(2,C)$ transformation as follows.

\vskip 0.2cm\noindent{\bf Proposition 4.}
Let $\kappa=\epsilon=0$, $\tau=\pi=0$, and $\bar\alpha+\beta=0$. Then, if $\rho\bar\mu-\bar\rho\mu=0,$ it is possible to set
\begin{eqnarray}
\gamma-\bar\gamma=\mu-\bar\mu. 
\end{eqnarray}

\noindent{\it Proof.}
The imaginary part of $\gamma$ transforms as 
\begin{eqnarray*}
\gamma'-\bar{\gamma}'&=& A^{-1}
\left(  \gamma -\bar{\gamma}
+i\Delta\theta\right),
\end{eqnarray*}
where $\gamma'$ is the value of $\gamma$ after the transformation. We look for the consistency of the equation 
\begin{eqnarray*}
A^{-1}
\left(  \gamma -\bar{\gamma}
+i\Delta\theta\right)&=& 
A^{-1}(\mu-\bar\mu). 
\end{eqnarray*}
When $D\theta=0$ \cite{Baysazan}, applying $D$ to the last equation, we obtain,
\begin{eqnarray*}
D\gamma -D\bar{\gamma}
+iD\Delta\theta&=& D\mu-D\bar\mu.
\end{eqnarray*}
 If we have $D\theta=0$, then $D\Delta\theta=0$, and the last equation yields
 \begin{eqnarray*}
   \Psi_2-\bar\Psi_2=\bar\rho\mu-\rho\bar\mu+\Psi_2-\bar\Psi_2,   
 \end{eqnarray*}
thus the integrability condition for $\theta$ is satisfied provided that  $\bar\rho\mu-\rho\bar\mu$.
\vskip 0.2cm

The expression of  $\gamma$  is
\begin{eqnarray}
\gamma=-\bar\mu+\frac{1}{2}\frac{\Psi_2-\bar\Psi_2}{\rho-\bar\rho},
\end{eqnarray}
and it is compatible with the NP equations involving $\gamma$.
\subsection{\bf Integrability of the NP system}

\vskip0.2cm\noindent{\bf Systems for $\rho$, $\bar\rho$ and $\mu $:}
We recall that the $\delta$ and $\bar\delta$ derivatives of $\rho$, $\bar\rho$ and $\mu$  are compatible provided that $\bar\mu$, $\Phi_{00}$ and $\Phi_{22}$ satisfy the algebraic relations given in the previous section. 
\begin{eqnarray}
&&\delta\rho=\bar\delta\rho=0,\quad  
\rho\Delta\rho+\mu D\rho=0,\\
&&\delta\bar\rho=\bar\delta\bar\rho=0,\quad  
\rho\Delta\bar \rho+\mu D\bar\rho=0,\\
&&\delta\mu=\bar\delta\mu=0,\quad  
\rho\Delta\mu+\mu D\mu=0.
\end{eqnarray}
It is checked that the systems for $\rho$, $\bar\rho$ and $\mu$  are integrable.
The $D$ derivatives of $\rho$, $\bar\rho$ and  $\mu$,
are compatible with their $\delta$, $\bar\delta$ and $\Delta$ derivatives.

\vskip0.2cm\noindent{\bf Systems for $\alpha$ and $\bar\alpha$: }
It is shown that  the system for $\alpha$, given by  
\begin{eqnarray}
&&D\alpha=\rho\alpha, \quad \Delta\alpha= -\mu\alpha,\quad
D\bar\alpha=\bar\rho\bar\alpha, \quad \Delta\bar\alpha= -\bar\mu\bar\alpha,\\
&&\delta\alpha+\bar\delta\bar\alpha-4\alpha\bar\alpha=
\mu\rho+\bar\mu\bar\rho-\mu\bar{\rho}
-\textstyle{\frac{1}{2}}(\Psi_2+\bar\Psi_2)+\Phi_{11}+\Lambda,
\end{eqnarray}
is compatible provided that 
\begin{eqnarray}
&&\rho\Delta\Lambda+\mu D\Lambda=0.\label{lambda}
\end{eqnarray}
As this is the only equation for $\Lambda$, there are no integrability conditions.

Note that as opposed to $\rho$ and $\mu$ that are fixed up to an arbitrary constant, NP equations determine $\alpha$ and $\bar\alpha$ up to arbitrary functions of two variables, subject to (II.36).  As discussed in \cite{Baysazan}, the left-hand side of (II.36) is the curvature 
$$K=\left[  \delta\alpha+\overline\delta\overline\alpha
-4\alpha\overline\alpha\right],$$
of the submanifolds with co-tangent frame $\{m,\bar m\}$, that are duals of the tangent vector fields $\delta$ and $\bar\delta$.
Using (II.36),  
\begin{eqnarray}
 K=\mu\rho+\bar\mu\bar\rho\!-\!\mu\bar{\rho}
-\frac{1}{2}(\Psi_2+\bar\Psi_2)+\Phi_{11}+\Lambda,
\end{eqnarray}
and it can be seen that $K$ satisfies the equations below. 
\begin{eqnarray}
&&\delta K=6\delta\Lambda,\quad \bar\delta K=6\bar\delta\Lambda,  \\
&& DK=(\rho+\bar\rho)K,\quad \Delta K=-(\mu/\rho) (\rho+\bar\rho) K.  
\end{eqnarray}

\vskip0.2cm\noindent{\bf Systems for $\Psi_2$, $\bar\Psi_2$ and $\Phi_{11}$: }
We  obtained the following system for $\delta$ and $\bar\delta$ derivatives, which is shown to be integrable
\begin{eqnarray}
&&\delta(\Psi_2+2\Lambda)=\bar\delta(\Psi_2+2\Lambda)=0,\;  \to \;
\rho\Delta(\Psi_2+2\Lambda)+\mu D(\Psi_2+2\Lambda)=0,\\
&&\delta(\bar\Psi_2+2\Lambda)=\bar\delta(\bar\Psi_2+2\Lambda)=0,\;  \to \;
\rho\Delta(\bar\Psi_2+2\Lambda)+ \mu D(\bar\Psi_2+2\Lambda)=0,\\
&&\delta(\Phi_{11}-3\Lambda)=\bar\delta(\Phi_{11}-3\Lambda)=0, \; \to \;
\rho\Delta(\Phi_{11}-3\Lambda)+\mu D(\Phi_{11}-3\Lambda)=0.
\end{eqnarray}
Derivatives of $\Psi_2$ and $\bar\Psi_2$  in $D$  and $\Delta$ directions are given by the Bianchi identities as below.
\begin{eqnarray}
&&D(\Psi_2-\Phi_{11}-\Lambda)
+\Psi_2\left[-3\rho+\frac{\rho\bar\rho}{\rho-\bar\rho}   \right]
-\bar\Psi_2\frac{\rho^2}{\rho-\bar\rho}  
+2\bar\rho\Phi_{11}-2\rho\Lambda=0,\\
&&D(\bar\Psi_2-\Phi_{11}-\Lambda)
+\bar\Psi_2\left[-3\bar\rho-\frac{\rho\bar\rho}{\rho-\bar\rho}   \right]
+\Psi_2\frac{\bar\rho^2}{\rho-\bar\rho}  
+2\rho\Phi_{11}-2\bar\rho\Lambda=0.
\end{eqnarray}
The compatibility of these two equations gives the second-order integrability conditions for $\Phi_{11}+3\Lambda$.
\begin{eqnarray}
&&\delta \Delta (\Phi_{11}+3\Lambda)+(\mu+\bar\mu)\delta(\Phi_{11}+3\Lambda)=0,\\
&&\bar \delta \Delta  (\Phi_{11}+3\Lambda)+(\mu+\bar\mu)\bar\delta(\Phi_{11}+3\Lambda)=0.
\end{eqnarray}
These new equations should be consistent with $\Delta(\Phi_{11}+3\Lambda)$ and with each other. This requires checking third-order commutators. These have been checked, and no new integrability conditions are obtained.

\vskip0.2cm\noindent
We can summarize the results as below.
\vskip0.2cm\noindent

{\bf Proposition 5. }
Let $(M,g)$  be a Petrov Type D spacetime where the repeated null directions of the Weyl tensor are geodesic and shear-free, the gauge is chosen to set $\Psi_i=0$  for $i\ne 2$, and the Ricci tensor has subspace spanned by the null directions and its complement are invariant subspaces of the Ricci tensor in mixed form, i.e,
$$\kappa=\sigma=\nu=\lambda=0,\quad  \Psi_0=\Psi_1=\Psi_3=\Psi_4=0, \quad  \Phi_{01}=\Phi_{12}=\Phi_{02}=0.$$
Then,
\begin{enumerate}
    \item[i.] $\kappa=\sigma=\nu=\lambda=0$, $\Psi_i=0$ for $i\ne2$, $\Phi_{01}=\Phi_{12}=\Phi_{02}=0$.
    \item [ii.] $\gamma$, $\bar\mu$, $\Phi_{00}$ and $\Phi_{22}$ are algebraically determined, 
    \item [iii.] $\{D\rho,\Delta\rho,\delta\rho,\bar\delta\rho\}$, 
    $\{D\bar\rho,\Delta\bar\rho,\delta\bar\rho,\bar\delta\bar\rho\}$ and 
    $\{D\mu,\Delta\mu,\delta\mu,\bar\delta\mu\}$ are determined,
    \item [iv.] $\{D\Psi_2,\Delta\Psi_2,\delta\Psi_2,\bar\delta\Psi_2\}$ and 
    $\{D\bar\Psi_2,\Delta\bar\Psi_2,\delta\bar\Psi_2,\bar\delta\bar\Psi_2\}$ are determined,
    \item [v.]$\{D(\Phi_{11}-3\Lambda),\delta(\Phi_{11}-3\Lambda), \bar\delta(\Phi_{11}-3\Lambda)  \}$ are determined,
    \item [vi.]$\{D(\Phi_{11}+3\Lambda),\delta\Delta(\Phi_{11}+3\Lambda), \bar\delta\Delta(\Phi_{11}+3\Lambda)  \}$ are determined after rearranging ($\ref{lambda}$).
\end{enumerate}
Thus the derivatives
\begin{eqnarray*}  
&&\{ \Delta\left(\Phi_{11}\!+\!3\Lambda\right),  \delta\left(\Phi_{11}+3\Lambda\right), \bar{\delta}\left(\Phi_{11}+3\Lambda\right),
\Delta\left(\Phi_{11}\!-\!3\Lambda\right)\} 
\end{eqnarray*}
are free.


\section{Admissible Sources} \label{sec3}
In this section, we will present the Segre types \cite{Stephani:2003tm} of admissible sources obtained in previous sections. Canonical forms of Segre types are given in \cite{McIntosh} and classification problems are discussed in \cite{Carminati}, but in the present work, as the Ricci tensor is block diagonal, we will be able to express its eigenvalues directly.

We note that $\Delta\phi+(\mu/\rho) D\phi=0$ is actually a symmetry. All variables satisfy this relation.  We assumed that $\Phi_{01}=\Phi_{12}=0$. As $\Phi_{00}$ and $\Phi_{22}$ are determined algebraically, sources are determined by $\Lambda$ and $\Phi_{11}$.
\begin{itemize}
    \item There are three equations for $\Phi_{11}$. Its evolution in $\Delta$ direction is free.
    \item All derivatives of $\Psi_2$ and $\bar\Psi_2$  are determined. Hence $\Psi_2$  depends on an arbitrary complex number.
    \item  $\Phi_{00}$ and $\Phi_{22}$ are algebraically determined.  
\item $\rho$, $\bar\rho$, $\mu$ are determined up to arbitrary constants.
\item The freedom in $\alpha$ is related to the curvature of the spacelike $2$-manifolds.
\item $\bar\mu$ and $\gamma$  are algebraically determined. 
\end{itemize}

The expression of the Ricci tensor in the mixed form is as below \cite{bilgedaghan}. 
\begin{eqnarray}
 R_{a}^b=2\left(%
\begin{array}{cccc}
  -(\Phi_{11}-3\Lambda) & -\Phi_{00} & \Phi_{10} & \Phi_{01} \\
-\Phi_{22}& -(\Phi_{11}-3\Lambda) & \Phi_{21} & \Phi_{12} \\
 - \Phi_{12} & -\Phi_{01} & (\Phi_{11}+3\Lambda) & \Phi_{02}\\
 - \Phi_{21} & -\Phi_{10} & \Phi_{20} & (\Phi_{11}+3\Lambda)\\
\end{array}%
\right).   
\end{eqnarray}
As $\Phi_{01}=\Phi_{12}=\Phi_{02}=0$, the eigenvalues are as follows.
\begin{eqnarray}
\lambda_1&=&
-2\Phi_{11}+6\Lambda-2\sqrt{\Phi_{00}\Phi_{22}}
\ (\text{timelike}),\\
\lambda_2&=&
-2\Phi_{11}+6\Lambda+2\sqrt{\Phi_{00}\Phi_{22}}
\ (\text{spacelike}),\\
\lambda_3&=&\lambda_4=2\Phi_{11}+6\Lambda \ (\text{spacelike}).
\end{eqnarray}
Using the relation $\mu\bar\rho=\bar\mu\rho$, it can be seen that $\Phi_{22}=\frac{\mu^2}{\rho^2}\Phi_{00}$, hence $\Phi_{00}\Phi_{22}$ is nonnegative and there are no complex eigenvalues.
So there is a repeated eigenvalue, and $\lambda_1$ and $\lambda_2$ are real.

Note that we are giving here the classification of the Ricci tensor with its trace, thus the case with all eigenvalues equal corresponds to a solution with the $\Lambda$ term only.  The Segre types and the freedom in the system are given below.
\begin{itemize}
    \item   [1)] Segre Type  $[1,1(11)]$: 
        $\lambda_1\ne \lambda_2\ne\lambda_3$,
    \item[2)] Segre Type  $[(1,1)(11)]$: $\lambda_1=\lambda_2$  
   with $\Phi_{00}=0$, 
     \item[3)] Segre Type  $[(1,111)]$: $\lambda_1=\lambda_2=\lambda_3$
      with $\Phi_{00}=\Phi_{11}=0$,
    \item[4)] Segre Type  $[(1,11)1]$: $\lambda_1=\lambda_3$
    with $2\Phi_{11}=-\sqrt{\Phi_{00}\Phi_{22}}$,
    \item[5)] Segre Type  $[1,(111)]$:
$\lambda_2=\lambda_3$
 with $2\Phi_{11}=\sqrt{\Phi_{00}\Phi_{22}}$.
\end{itemize}

\subsection{Examples}

\vskip0.2cm\noindent
{\bf  Case 1: Segre Type $[1,1(11)]$ ($\lambda_1\ne \lambda_2\ne\lambda_3$):} Anisotropic perfect fluid solutions \cite{McIntosh,Carminati}. 
Without any additional assumption, the Segre type is $[1,1(11)]$. 
 The energy-momentum tensor of the fluid is
\begin{equation}
T_{\mu\nu}\!=\!diag(-\rho,p_r,p_t,p_t),
\end{equation}
where $\rho$ is energy density, $p_r$, $p_t$ are radial and tangential pressures. A solution in this Segre type is presented in Example 1 of Appendix \ref{example} \cite{Ellis1967,Stewart1968}. Note that the NP quantities for this example of the given Segre type are consistent with our results.

In this type, we have $\Phi_{00}=\Phi_{22}\neq0$ and $\Phi_{11}\neq0$.

\vskip0.2cm\noindent
{\bf  Case 2: Segre Type  $[(1,1) (11)]$ ($\lambda_1=\lambda_2$  
   with $\Phi_{00}=0$):}  
  \vskip0.2cm\noindent

We consider two cases, $\Lambda=0$ and $\Lambda\ne 0$.
   
    In the case $\Lambda=0$, $\Phi_{00}=0$ yields 
\begin{eqnarray*}
 -\Psi_2\bar\rho+\bar\Psi_2\rho+2\rho\bar\rho(\mu-\bar\mu)=0.   
\end{eqnarray*}
$\delta$ and $\bar\delta$  derivatives of $\Phi_{11}$ are identically zero and its $D$ and $\Delta$ derivatives give,
\begin{eqnarray*}
&&D\Phi_{11}=2(\rho+\bar\rho)\Phi_{11},\quad
\Delta\Phi_{11}=-2(\mu+\bar\mu)\Phi_{11},
\end{eqnarray*}
which corresponds to the  Einstein-Maxwell solution,
\begin{eqnarray*}
\Phi_{11}=\phi_1\bar\phi_1,\quad D\phi_1=2\rho\phi_1,\quad 
\Delta\phi_1=-2\mu\phi_1,\quad \delta\phi_1=\bar\delta\phi_1=0.
\end{eqnarray*}
For this group, the well-known Petrov type D example is the electromagnetic Reissner--Nordstrom black hole solution \cite{Prasanna1968},
\begin{eqnarray}
\hspace{-1cm}ds^2 = \left( -\frac{Q^{2} - 2 \, M r + r^{2}}{r^{2}} \right) \mathrm{d} t^2 + \left( \frac{r^{2}}{Q^{2} - 2 \, M r + r^{2}} \right) \mathrm{d} r^2 + r^{2} \mathrm{d} {\theta}^2 + r^{2} \sin\left({\theta}\right)^{2} \mathrm{d} {\phi}^2,
\end{eqnarray}
where $M$ is the mass and $Q$ is the electric charge parameter of the black hole.

The second group of solutions corresponds to the case 
 $\Phi_{00}=\Phi_{22}=0$ but  $\Lambda\ne 0$.  $D$ and $\Delta$ derivatives read
\begin{eqnarray*}
&&D(\Phi_{11}+3\Lambda)=2(\rho+\bar\rho)\Phi_{11},\quad
\Delta(\Phi_{11}+3\Lambda)=-2(\mu+\bar\mu)\Phi_{11}.
\end{eqnarray*}

In this case, under our primary geometric assumptions, specifically the requirement that $\Phi_{01}=\Phi_{12}=0$ alongside the conditions for twisting ($\rho \neq \bar{\rho}$) and expanding ($\rho + \bar{\rho} \neq 0$) null congruences, the resulting system of NP equations is found to be formally consistent. The Riquier-Janet integrability analysis, performed via the REDUCE computer algebra system, confirms that the field equations remain involutive without generating further constraints that would force a reduction to vacuum or non-twisting solutions.

However, a crucial distinction must be made regarding existing solutions in the literature that share this Segre classification. For instance, the well-known Bertotti-Robinson metric \cite{Kramer1978} is often cited as a representative of this algebraic class. Nevertheless, it is fundamentally incompatible with the framework developed here for two decisive reasons:
\begin{enumerate}
	\item {Kinematical Incompatibility:} The Bertotti-Robinson solution is non-twisting ($\rho = \bar{\rho}$), whereas our setup is strictly defined on a twisting NUT background.
	\item {Geometric Misalignment:} The specific configuration of the Ricci components in the Bertotti-Robinson geometry does not satisfy the restricted set of geometric assumptions ($\Phi_{00},\Phi_{11},\Phi_{22}\neq0$) imposed in our study.
\end{enumerate}

Consequently, while Case 2 represents a mathematically consistent case where integrability is preserved, this case does not satisfy the restricted set of geometric assumptions($\Phi_{00},\Phi_{11},\Phi_{22}\neq0$) imposed in our study.

 
\vskip0.2cm\noindent
{\bf  Case 3: Segre Type  $[(1,111)]$($\lambda_1=\lambda_2=\lambda_3$
      with $\Phi_{00}=\Phi_{11}=0$):}  This corresponds to vacuum metrics \cite{McIntosh,Carminati}.  The most well-known Type D solutions representing this case are the Schwarzschild and Kerr metrics. 
      
      The Schwarzschild spacetime \cite{Droste1917,Reissner1916} is
\begin{eqnarray}
ds^{2}
= -\left(1-\frac{2M}{r}\right) dt^{2}
  + \left(1-\frac{2M}{r}\right)^{-1} dr^{2}
  + r^{2}\left(d\theta^{2}+\sin^{2}\theta\, d\phi^{2}\right),
\end{eqnarray}

and the Kerr spacetime \cite{Kerr1979} is
\begin{eqnarray}
ds^{2}=&& -\left(1 - \frac{2Mr}{\Sigma}\right) dt^{2}
        - \frac{4Mar\sin^{2}\theta}{\Sigma}\, dt\, d\phi
        + \frac{\Sigma}{\Delta}\, dr^{2}
        + \Sigma\, d\theta^{2} \nonumber \\
&&+\left(r^{2}+a^{2} + \frac{2Ma^{2} r \sin^{2}\theta}{\Sigma}\right)
          \sin^{2}\theta\, d\phi^{2},
\end{eqnarray}
with
\begin{eqnarray}
\Sigma = r^{2} + a^{2}\cos^{2}\theta, \quad
\Delta = r^{2} - 2Mr + a^{2}.
\end{eqnarray}
The Schwarzschild spacetime is a non-twisting Petrov type~D vacuum solution, whereas the Kerr spacetime is a twisting Petrov type~D vacuum solution. 

However, neither of these spacetimes is compatible with our results. The exclusion does not arise solely from kinematical considerations, but rather from the fundamental geometric assumptions imposed in our construction. In particular, our analysis is restricted to non-vacuum solutions with non-vanishing Ricci components, whereas the condition
\begin{equation}
	\Phi_{ij}=0, \quad i,j=0,1,2,
\end{equation}
enforces a strictly vacuum structure. Consequently, Case~3 lies entirely outside the scope of our NUT-background, non-vacuum class of solutions.
\vskip0.2cm\noindent
{\bf  Case 4: Segre Type $[(1,11)1]$ ($\lambda_1=\lambda_3$
    with $2\Phi_{11}=-\sqrt{\Phi_{00}\Phi_{22}}$):} Tachyon fluid solutions \cite{Carminati,McIntosh} with $2\Phi_{11}=-\sqrt{\Phi_{00}\Phi_{22}}$. We represent the energy-momentum tensor of the fluid in the following form,
\begin{eqnarray}
{T^{\mu}}_{\nu}\!=\!diag\Big[-\rho,p_x,p_y,p_z\Big].
\end{eqnarray}
The most well-known solutions representing this case are Bianchi type-I spacetimes \cite{Taub1951}.
 Since this metric is of Petrov type~I, it is therefore incompatible with our result.However, such spacetimes are generically of Petrov type~I, indicating a fully non-degenerate Weyl tensor structure. This is in direct contrast with the algebraically special Petrov type D considered in our framework. Consequently, despite satisfying the required Ricci algebraic condition, these solutions are excluded due to their incompatible Weyl classification. Therefore, Case~4 does not yield admissible solutions within our NUT-background, algebraically special setting which is defined by assumptions on the spin coefficents.

\vskip0.2cm\noindent
{\bf Case 5: Segre type  $[1,(111)]$($\lambda_2=\lambda_3$
 with $2\Phi_{11}=\sqrt{\Phi_{00}\Phi_{22}}$):} Perfect fluid solutions \cite{McIntosh,Carminati}. 
The energy-momentum tensor of a perfect fluid is
\begin{eqnarray}
 T_{\mu\nu}=(\rho+p)u_{\mu}u_{\nu}+pg_{\mu\nu}. \end{eqnarray}
As $\Phi_{02}=0$, we should have $p=0$, hence the dust solutions are characterized by the triple eigenvalue being zero. This corresponds to the case
\begin{eqnarray}
 \sqrt{\Phi_{00}\Phi_{22}}=2\Phi_{11},\quad \Phi_{11}=-\Lambda.   
\end{eqnarray}
It also gives the topological vacuum case. A solution in this Segre type is presented in Example 2 of Appendix \ref{example} \cite{Godel1949}. The NP quantities for this example of the given Segre type are consistent with our results.




\section{CONCLUSIONS}\label{sec6}
We studied shear-free, expanding and twisting non-vacuum Petrov Type D solutions on the Newman–Unti–Tamburino (NUT) background. 

First, we assumed that the spacetime is non-vacuum and Petrov Type D; hence, we kept only the non-zero Weyl component $\Psi_{2}$. We then investigated expanding and twisting metrics, so that the Newman–Penrose spin coefficients satisfy $\kappa=\sigma=\lambda=\nu=0$, while $\rho$ and $\mu$ are non-zero and complex.  We set $\epsilon\!=\!0$  as the gauge condition using the $SL(2,C)$ rotations of type~B. We analyzed the Newman-Penrose equations in which the spin coefficients $\rho$ and $\mu$ are the non-zero complex cases. For the $(\rho-\bar{\rho}\neq0,\;\mu-\bar{\mu}\neq0)$, corresponding to complex values of the spin coefficients $\rho$ and $\mu$, we obtained the integrability conditions of the Newman-Penrose equations. In our calculations, we obtained the well-known algebraic relation $\rho\ \bar{\mu} - \mu\ \bar{\rho} = 0.$ Then we algebraically determined $\gamma,\bar{\mu},\Phi_{00}$ and 
$\Phi_{22}$. To ensure the integrability of the system, we determined the directional derivatives of $\rho$, $\bar{\rho}$, $\mu$, $\Psi_2+2\Lambda$, and $\bar{\Psi}_2+2\Lambda$. We showed that the conditions determine the solutions up to a freedom in $\Phi_{11}\pm3\Lambda$. We showed that the freedom in $\alpha$ is related to the curvature of the 2-spaces and obtained the directional derivatives of the curvature of space-like submanifolds in terms of the spin coefficients.

We investigated the Ricci classification, which gives a canonical list of the trace-free Ricci tensor components for each Plebanski class in terms of the tetrad formalism of Newman and Penrose with non-zero Ricci components $\Phi_{00}, \Phi_{11}$ and $\Phi_{22}$. We found that when $\Phi_{00}, \Phi_{11}$ and $\Phi_{22}$ are all non-zero (Case\;$1$), the system corresponds to anisotropic perfect fluid solutions. When $\Phi_{00} = \Phi_{22} = 0$ and only $\Phi_{11}$ is non-zero (Case\;$2$), the system corresponds to an Einstein-Maxwell solution. Furthermore, when these Ricci components become zero $(\text{Case}\;3)$, the system reduces to the vacuum case. Under the condition $2\Phi_{11} = -(\mu / \rho)\,\Phi_{00}$ $(\text{Case}\;4)$, the system corresponds to a tachyon fluid solution. Under the condition $2\Phi_{11} = (\mu / \rho)\,\Phi_{00}$ $(\text{Case}\;5)$, the system corresponds to a perfect fluid solution. 
 
We found that the Newman-Penrose quantities obtained for the Gödel spacetime and locally rotationally symmetric spacetimes are consistent with our results in this work.

These results provide an understanding of the algebraic and physical properties of exact solutions in a coordinate-free manner using the Newman–Penrose formalism for shear-free, expanding, and twisting non-vacuum Petrov Type D spacetimes.

All computations in this work including the checking of the integrability conditions of the Newman-Penrose equations, the verification of the  involution of the Riquier--Janet system, and the checking of third-order commutator relations were carried out using the {\sc Reduce} computer algebra system. This allowed us to treat the higher-order compatibility conditions in a systematic and reliable manner. The approach lends itself directly to obtain directional derivatives of spin coefficients, the verification of gauge transformations, and the systematic checking of Bianchi identities, suggesting that the methods developed here may serve as a practical foundation for computer-assisted searches for exact solutions in more
general geometric settings.
\begin{acknowledgments}
This work is funded by TUBITAK 1001 Program, Grant Number 123R114.
\end{acknowledgments}
\begin{appendix}
\section{Example Metrics Consistent with Our Results}
\label{example}
\vskip 0.2cm\noindent
{\bf Example 1 (Locally Rotationally Symmetric Metrics):}
The metric in local coordinates $(t,x,y,z)$ is given by
\begin{eqnarray}
  ds^{2}= a\left[- dt^{2}+ A^{2}(dx+\phi \ dz)^2) \right]
+ B^{2}\left[
dy^{2} + (\phi_{y} dz)^2 \right],  
\end{eqnarray}
where  $a=\pm 1$, $A=A(t)$, $B=B(t)$, $\phi=\phi(y)$,   $k$ is the curvature constant that takes values $0,\pm 1$ with 
$$\phi(y,1)=\cos(y),\quad \phi(y,0)=y^2/2, \quad \phi(y,-1)=\cosh(y).$$
The tetrad is chosen as 
$$
l=\frac{\sqrt{a}}{\sqrt{2}}\left[ dt+A (dx+\phi dz)\right],\quad
n=\frac{\sqrt{a}}{\sqrt{2}}\left[ dt-A (dx+\phi dz)\right],\quad
m = \frac{B}{\sqrt{2}} \left[ dy + i \phi_{y} dz \right].
$$
The nonzero NP quantities for this metric are given by
\[
\rho\!=\!-\!\mu\!=\!
= \frac{1}{\sqrt{2a}}
\left[
\frac{B_t}{B}
+ i\,\frac{aA}{2B^{2}}
\right],
\qquad\alpha\!=\!-\beta\!=\!
 \frac{\phi}{2\sqrt{2}\,B\,\phi_{y}}
\;,
\]
\[
\gamma\!=\!-\epsilon
= \frac{1}{2\sqrt{2a}}
\left[
\frac{ A_t}{A}
- i\,\frac{aA}{2B^{2}}
\right]\;,
\]
\[
\Phi_{00}
= \Phi_{22}
= -\frac{1}{2a}
\left[
\frac{ B_{tt}}{B}
- \frac{ A_{t}  B_{t}}{AB}
- \frac{a^{2}A^{2}}{4B^{4}}
\right]\;,
\]
\[
\Phi_{11}
= -\frac{1}{4a}
\left[
\frac{A_{tt}}{A}
- 2\frac{ B_t}{B}
- \frac{a}{B^{2}}
+ \frac{3}{4}\frac{a^{2}A^{2}}{B^{4}}
\right]\;,
\]
\[
\Psi_{2}
= -\frac{1}{6a}
\left[
\frac{ A_{tt}}{A}
- \frac{B_{tt}}{B}
- \frac{A_{t} B_{t}}{AB}
+ 2\frac{B_{t}}{B}
- \frac{a^{2}A^{2}}{B^{4}}
+ \frac{a}{B^{2}}
+ \frac{3ia}{B^{2}}(A\frac{ B_t}{B} -  A_t)
\right]\;,
\]
\[
\Lambda
= -\frac{1}{4a}
\left[
- \frac{A_{tt}}{3A}
- \frac{2}{3}\frac{ B_{tt}}{B}
- \frac{2}{3}\frac{A_{t} B_{t}}{AB}
- \frac{2}{3}\frac{B_t}{B}
- \frac{a}{3B^{2}}
+ \frac{a^{2}A^{2}}{12B^{4}}
\right]\;.
\]

Here, subscripts $t$ and $y$ are used to denote  derivatives,e.g., $A_t = d A / d t$ and $\phi_y = d \phi / d y$. Higher-order derivatives are denoted similarly (e.g., $\phi_{yy} = d^2 \phi / d y^2$). 

\vskip 0.2cm\noindent
{\bf Example 2 (The Gödel Spacetime):}
The Gödel metric in local coordinates $(t,x,y,z)$ is given by
 \begin{eqnarray}
 ds^2=-a^2 \Big[-\left(dt+e^xdy\right)^2+dx^2+\frac{1}{2}e^{2x}dy^2+dz^2\Big],    
 \end{eqnarray}
where $a$ is a constant.  The tetrad is chosen as
\begin{eqnarray*}
 &&   l=\frac{a}{\sqrt{2}}\left[ dt+e^x dy + dz\right],\quad
    n=\frac{a}{\sqrt{2}}\left[ dt+e^x dy -dz\right],\quad
    m=\frac{a}{\sqrt{2}}\left[  dx+\frac{i}{\sqrt{2}}e^xdy\right].
\end{eqnarray*}
The nonzero NP quantities are  
\begin{eqnarray*}
&&\rho=\mu=-\frac{i}{2a},\quad \alpha=-\frac{\sqrt{2}}{4a}, \quad  \epsilon=\gamma=-\frac{i}{4a},\\
&&\Phi_{00}=\Phi_{22}=2\Phi_{11}=\frac{1}{4a^2},\quad \Phi_{11}+3\Lambda=0,\quad 
\Phi_{11}-3\Lambda=\frac{1}{4a^2},\\
&&\Psi_2=-\frac{1}{6a^2}.
\end{eqnarray*}

This solution describes a rotating, homogeneous, non-expanding universe filled with pressureless dust and a cosmological constant. The parameter ``$a$" determines the magnitude of the rotation of the universe: a larger ``$a$" corresponds to a slower rotation and smaller matter density. Due to the cross term, the spacetime is stationary but not static. It corresponds to the global rotation of the universe. The Ricci components satisfy $\sqrt{\Phi_{00}\Phi_{22}}=2\Phi_{11}$.

\end{appendix}
\bibliography{references}   

\end{document}